\documentclass[aps,pra,preprint,groupeaddress,amsmath,amssymb,showpacs,preprintnumbers]{revtex4}
\usepackage{graphicx}
\usepackage{dcolumn}
\usepackage{bm}
\input epsf

\begin{document}
\preprint{}
\title{Candidate molecular ions for an electron electric dipole moment experiment}

\author{Edmund R. Meyer}
\email{meyere@murphy.colorado.edu}
\affiliation{JILA, National Institute of Standards and Technology
and University of Colorado, Department of Physics,
Boulder, Colorado 80309-0440, USA}
\author{Michael P. Deskevich}
\affiliation{JILA, National Institute of Standards and Technology
and University of Colorado, Department of Chemistry and Biochemistry,
Boulder, Colorado 80309-0440, USA}
\author{John L. Bohn}
\affiliation{JILA, National Institute of Standards and Technology
and University of Colorado, Department of Physics,
Boulder, Colorado 80309-0440, USA}

\date{\today}

\begin{abstract}
This paper is a theoretical work in support of a newly proposed 
experiment (R. Stutz and E. Cornell, Bull. Am. Soc. Phys. {\bf 89}, 76 2004) 
that promises greater sensitivity to measurements of
the electron's electric dipole moment (EDM) based on the trapping of 
molecular ions.  Such an experiment requires the choice
of a suitable molecule that is both experimentally feasible and
possesses an expectation of a reasonable EDM signal.  We find that
the molecular ions PtH$^+$, HfH$^+$, and HfF$^+$ are suitable
candidates in their low-lying $^3\Delta$ states.  
In particular, we anticipate that the effective electric fields
generated inside these molecules are approximately of 73 GV/cm, 
-17 GV/cm, and -18 GV/cm respectively.
As a byproduct of
this discussion, we also explain how to make  estimates of the
size of the effective electric field acting in a molecule, using
commercially available {\it nonrelativistic} molecular structure
software.
\end{abstract}


\maketitle
\section{Introduction}

The Standard Model of elementary particle physics admits that the electron
may have an electric dipole moment (EDM), but a darn small one;
the Standard Model prediction is on the order of $d_e = 10^{-38}$
e-cm, many orders of magnitude below the current experimental limit.
Nevertheless, in models beyond the Standard Model, notably super-symmetry,
a larger EDM is expected, sometimes tantalizingly close
to the current experimental upper bounds on $d_e$ \cite{barr}.  To observe $d_e$
would imply a different form of charge-conjugation/parity (CP) violation
than the familiar form in the K- and B-meson systems. The observation of $d_e$ 
requires extremely large electric fields to observe the energy shift due to 
different alignments of $d_e$ along the quantization axis. 

The current upper 
limit on the electron EDM,$1.6 \times 10^{-27}$ e-cm, comes from measurements in atomic 
thallium \cite{Commins}.  In this
experiment, an {\it effective} electric field is generated inside the
atom.  The magnitude of this electric field (on the order of
70 MV/cm) is far larger
than the field that could be applied directly in the laboratory. This 
circumstance is what makes an atomic EDM experiment desirable.

Taking this idea even further, Sandars noted that the effective 
electric field within a molecule
can be even larger than that within an atom \cite{Sandars}.  This insight
has prompted a series of proposed and actual experiments, aimed at 
using molecules as high-electric-field laboratories.  These experiments have
employed either molecular beams or traps for neutral molecules, including 
molecules PbO \cite{PbO,PbO2} and YbF \cite{YbF,YbF2}. Several others have 
also been proposed as viable candidates \cite{TlF1,TlF2,PbF,BaF,Derevianko,Isaev}.  

Recently, a new experiment has been proposed that may increase the 
experimental sensitivity to an electron EDM by  orders of magnitude \cite{Cornell,c2}.  The
key to this experiment is using molecular ions in lieu of neutral
molecules.  Because ions are far easier to trap for long times than
neutrals, the proposed experiment wins in terms of long coherence times.
Nevertheless, such an experiment poses a number of challenges, both in experimental
technology and in the choice of a reasonable molecule.

The purpose of this paper is to identify diatomic molecular ions that
possess  properties that are desirable for this experiment. For instance, 
the molecule should have a relatively
small degeneracy of quantum mechanical states near its ground state.  This
feature, in a cold sample at thermal equilibrium, would ensure that most or all of the
ions are already in or near a state that is useful for the EDM measurement.
Because the number of ions in a trap is fairly small, this requirement
is necessary for achieving a good signal-to-noise ratio.  To meet this requirement
we restrict our attention to diatomic hydrides, where the
light hydrogen atom will contribute to a relatively large rotational
constant.  Similarly, we are interested in heavy atoms with spinless isotopes in order 
to ensure a favorable population distribution. This requirement is an 
experimental one. On the theory side, isotopes with spin are useful in characterizing the 
accuracy of the calculations and for making empirical estimates \cite{Isaev,PbO}.
A second requirement is that the electric field used to polarize the molecule 
must be relatively small, on the order of 100 V/cm or less. This is because 
in the proposed experiment, the polarizing field must be rotating
at MHz frequencies in order to act on the molecules without
accelerating them out of the trap. Lastly, we restrict ourselves to 
molecules that have at least one unpaired spin (paramagnetic, i.e. molecules 
with $S\ne 0$), so that the net EDM of the electrons does not vanish

Because we confine our attention to diatomic molecules, 
we must meet two additional requirements: (1) A large dipole moment,
on the order of 1 Debye, which is
not hard to achieve if the two atoms are quite different, and (2) A small 
energy splitting between even- and odd-parity branches
of the ground state.  In other words, a small omega doublet is 
desirable. 

In general, the size of the omega doublet
is smaller for larger quantum number $\Omega$, which represents the
total (orbital plus spin) angular momentum of the electron about
the intermolecular axis.  For this reason we prefer diatomic molecules with
high angular momentum.  In the following we consider species with
$^3\Delta$ molecular symmetry, which can support values of $\Omega$
as large as $\Omega=3$.  This choice is in sharp contrast with 
previous EDM candidate molecules, most of which were based on 
$^2\Sigma$ molecular symmetries.  More recently, molecular ions
of $^3 \Sigma$ \cite{PbO,PbO2} or $^2\Pi$ \cite{Derevianko,Isaev}
symmetry have been considered as well.

Finally, the essential requirement for the molecule is that it must possess
a large internal effective electric field.  It is this field that
acts on the electron's EDM to make a measurable signal.  A quick
review of the relevant physics reveals why a large internal field is so important.
If an electric field were applied to
an atom, and if relativity did not matter, then a given electron would
feel no net electric field.  In equilibrium, the applied external field
would be exactly balanced by the field due to all the other charges
in the atom, or else the electron would be pulled to a different
equilibrium.  This is the content of Schiff's theorem \cite{Schiff}.
However, the fact that the electron can move at relativistic speeds in
an atom provides a loophole around this theorem.  In the relativistic case,
the electron's equilibrium is governed by the balance of electric
fields, plus the motional magnetic field generated when it swings
by the nucleus.  Thus the net electric field, by itself, can be
non vanishing. By applying an external electric field, the non zero 
field within the molecule can be increased by a relativistic enhancement 
factor manifested through the mixing of $s$
and $p$ atomic states.  The resulting energy shift scales with
the third power of the nuclear charge, and therefore strongly favors 
heavy atoms as EDM candidates.  

Qualitatively, molecules can be even more effective at mixing $s$ and $p$
states of atoms than externally applied fields. Indeed, 
in molecular orbitals it is possible for $s$
and $p$ levels to perfectly hybridize, as in the conjugated bonds of
carbon-bearing molecules. The relativistic enhancement factor can be estimated
from its value for the heavy atom, therefore estimating the effective internal
electric field becomes a question for nonrelativistic molecular
structure theory.  
The mixing of $s$ and $p$ atomic orbitals is useful only when 
it produces a molecular orbital of $\sigma$ symmetry.  
An electron in such a molecular orbital has no orbital
angular momentum about the 
molecular axis, and can thus penetrate close to the heavy atom's
nucleus, where relativistic effects are greatest.  This behavior explains 
why molecules with total symmetry $\Sigma$ have been preferred in the past.  

As noted above, however, the experiment under consideration is
likely constrained to molecules with 
{\it nonzero} angular momentum about the axis.  To accomplish these
seemingly contradictory goals, we
seek molecules having two valence electrons.  One, of $\sigma$
molecular symmetry, will be responsible for the effective electric
field and consequent EDM signal.  The second, of $\delta$ symmetry,
will allow for small omega doubling.  We are thus led to contemplate
molecules of $(\sigma \delta)^3\Delta_{\Omega}$ symmetry overall.
Further, since $\delta$ molecular orbitals are likely to arise from
$d$-type atomic orbitals, we seek diatomic molecules in which the heavy
atom is a transition metal.  Previous theoretical work on heavy
transition metal hydrides \cite{Ohanessian} identifies a pair of
likely molecules with low-lying $^3 \Delta$ states, namely,
HfH$^+$ and PtH$^+$.  

In the following sections we will explore the properties of these and other
molecules in terms of their usefulness to an electron EDM experiment.  
In Sec.~\ref{sec2} we revisit the theoretical methods for estimating
the EDM signal in terms of nonrelativistic molecular wave functions 
and the way in which such wave functions can be constructed. 
In Sec.~\ref{sec3} we present concrete examples of calculations
for various diatomic molecules.  These calculations verify
that the perturbative methods we employ can semi quantitatively reproduce the
effective electric fields predicted previously by more substantial
calculations.

\section{Background: observable characteristics of the electron EDM}
\label{sec2}

\subsection{The effective electric dipole moment of an atom}

In EDM circles, it is customary to speak of an ``effective electric field''
experienced by the electron inside an atom or molecule.  
What this means in practice is that the energy shift 
due to an electron EDM of magnitude $d_e$ is simply
\begin{equation}
\Delta E = -d_e F_{\rm eff}.
\end{equation}
We use the letter $F$ to denote electric field strengths,
to distinguish them from energies, denoted $E$.
To evaluate the energy shift,
one writes the perturbation Hamiltonian in Dirac notation
(\cite{Khriplovich}, p. 254 ff.) is:
\begin{equation}
\label{EDM_Ham}
H_d = \left( \begin{array}{cc}
0 & 0 \\
0 & 2d_e {\bf \sigma} \cdot {\bf F} \\
\end{array} \right).
\end{equation}
This $2 \times 2$ matrix acts on the state vector $\psi = (\psi_U, \psi_L)$, 
where $\psi_U$ and $\psi_L$ are the ``upper'' and ``lower'' components of the 
electron's Dirac wave function.  Each of $\psi_U$ and $\psi_L$, in turn, stands for
a two-component  spinor.

In Eq.(\ref{EDM_Ham}), ${\bf \sigma}$ is a
Pauli matrix designating the electron's spin, so the combination 
$d_e {\bf \sigma}$ describes
the EDM of the electron.  (The direction of the electron's EDM 
is presumed to be collinear with its magnetic moment).
This moment interacts with the instantaneous electric field ${\bf F}$
experienced by the electron.  To a good approximation, this field is
given by the Coulomb field of the heavy nucleus with charge $Ze$, 
${\bf F} = Ze {\hat r}/r^2$.  Here $r$ is the electron-heavy atom
distance.  Here and throughout this paper we use atomic units,
setting $\hbar = m_e = 1$.

Next we denote the lower component of the electron's ground state
wave function to be $|\psi^0_L \rangle$.  Then the perturbative
energy shift due to $H_d$ would be 
$\langle \psi^0_L | 2d_e {\bf \sigma}\cdot {\bf F} | \psi^0_L \rangle$. 
However, this expression vanishes for 
a ground-state wave function of definite parity, owing to the
odd-parity of ${\bf F}$.  Thus, an externally applied electric
field is required to mix states of different parities.  In the
important example of an atom with an s-electron (e.g., Cs),
an applied electric field would create a linear combination
of $s$ and $p$ states,
\begin{equation}
|\psi^0_L \rangle = \epsilon_s |\psi_L(s) \rangle
+ \epsilon_p |\psi_L (p) \rangle.
\end{equation}
Higher angular momentum states are also possible in principle 
but do not make a substantial contribution at laboratory fields.
In typical atomic EDM experiments, the field is sufficiently weak
to invoke the perturbative approximations:
\begin{eqnarray}
\epsilon_s & \approx & 1 \\
\epsilon_p & \approx & \frac{\langle s | e{\bf r} \cdot {\bf F}_{\rm applied}
 | p \rangle}{E(s) - E(p)}.
\end{eqnarray}
Here $e{\bf r}$ stands for the relevant dipole operator, and the denominator
denotes the energy difference between the $s$ and $p$ states.

With these approximations, the expression for the energy shift
becomes
\begin{equation}
\label{EDM_shift}
\langle \psi^0_L | 2d_e{\bf \sigma}\cdot{\bf F} | \psi^0_L \rangle =
2 \epsilon_s \epsilon_p \left( \frac{d_e Ze}{a_0^2}\right) 
\Gamma_{\rm rel}.
\end{equation}
The factor $\Gamma_{\rm rel}$ incorporates the relativistic
effects and can be estimated analytically using relativistic
Coulomb wave functions \cite{Khriplovich}:
\begin{equation}
\Gamma_{\rm rel} = a_0^2 \langle s_{1/2} | \frac{2 \sigma \cdot {\hat r}}{r^2} 
| p_{1/2} \rangle =-\frac{4 (Z \alpha )^2 Z_i^2}
{\gamma (4 \gamma^2 - 1)  (\nu_s \nu_p)^{3/2}},
\end{equation}
where $\alpha$ is the fine structure constant and
$\gamma = \sqrt{ (j+1/2)^2-(Z \alpha)^2 }$. The quantities $\nu_s$ and
$\nu_p$ are the effective quantum numbers for the $s$ and $p$ states of the
heavy atom, defined as the the principal quantum number shifted by
a quantum defect. All values of $\nu_s$ and $\nu_p$ are obtained from Ref.~\cite{m58}.  
The constant $Z_i$ is the effective nuclear charge
seen by the valence electron; for a neutral atom $Z_i=1$.
An additional energy shift would arise from perturbations due to
the $p_{3/2}$ excited state.  However,
the matrix element connecting the $s_{1/2}$ state
to the $p_{3/2}$ state is vanishingly small,  since
the $p_{3/2}$ radial function has negligible amplitude near the
nucleus \cite{Kozlov}.

The energy shift of an atom in an electric field, due to 
the EDM of the electron thus requires 
mixing  the $s_{1/2}$ ground state and the $p_{1/2}$ 
excited state.  The resulting energy shift is, in atomic units,
\begin{equation}
\label{atom_shift}
\Delta E = \left[ \frac{4 \langle s_{1/2} | z | p_{1/2} \rangle}
{E(s_{1/2}) - E(p_{1/2})}
\Gamma_{\rm rel} \frac{Ze^2}{a_0^2}\right] d_e F_{\rm applied}.
\end{equation}
This expression can be viewed in two ways.  One way is to envision that
the atom gains an EDM that is bigger than the electron's EDM
by the factor in square brackets in Eq.(\ref{atom_shift}).  Alternatively,
the existing electronic EDM may be seen as experiencing the applied
field $F_{\rm applied}$, enhanced by this same factor.

\subsection{The effective electric field inside a molecule}

In diatomic molecules, the same kind of $s$-$p$ mixing may occur,
perhaps much more substantially than in an atom \cite{Flambaum,Khriplovich2}. 
If so, the basic expression for
the measured energy shift is still given by Eq.(\ref{EDM_shift}),
but the coefficients may have quite different values.
To estimate these coefficients, we expand the molecular orbital
wave function of the contributing electron in terms of
atomic orbitals:
\begin{equation}
|{\rm mol} \rangle = \epsilon_s |s \rangle + \epsilon_p |p \rangle
+ \sum_{\rm other} \epsilon_{\rm other} |{\rm other }\rangle.
\end{equation}
Here $s$ and $p$ still refer to the relevant 
atomic orbitals on the heavy atom, from which the EDM shift will arise.
But now there are other orbitals that may participate.  For example,
the electron wave function may have considerable contributions from 
$d$ or $f$ orbitals on the heavy atom or from atomic orbitals
residing on the other atom.  Thus, in general, it may {\it not} be
a good approximation that $\epsilon_s^2 + \epsilon_p^2=1$.
Nevertheless, one may seek molecules for which $|\epsilon_s| \approx
|\epsilon_p| \approx 1/\sqrt{2}$, which would be the optimal mixing for 
our purposes.  To determine these quantities for a given molecule
is a problem in nonrelativistic molecular orbital theory.

Once the amplitudes $\epsilon_s$ and $\epsilon_p$ are known,
we can use Eq.(\ref{EDM_shift}) to find the energy shift,
with the following caveat.  The orbital $|p \rangle$ here stands for
the nonrelativistic $p$ wave function of the atom.  It is therefore
a linear combination of the relativistic atomic states, denoted
by total spin $j$:
\begin{equation}
| p \rangle = - \frac{2 \sigma}{\sqrt{3}}|p_{1/2} \rangle
+ \sqrt{\frac{2}{3}} | p_{3/2} \rangle.
\end{equation}
The factor $2\sigma$ emphasizes that the $p_{1/2}$ contributes with
 opposite signs for the two spin projections on the molecular axis, 
$\sigma = \pm 1/2$.
As noted above, the $p_{3/2}$ orbital does not contribute to
the electron EDM shift; hence the relevant matrix element reads
\begin{equation}
\langle s | 2d_e{\bf \sigma}\cdot{\bf F} | p \rangle = -\frac{2 \sigma}{\sqrt{3}} 
\langle s_{1/2} | 2d_e{\bf \sigma}\cdot{\bf F} | p_{1/2} \rangle .
\end{equation}
Thus, following Flambaum \cite{Flambaum} and Khriplovich and 
Lamoreaux \cite{Khriplovich2}, we find the electron EDM energy shift
for a diatomic molecule (in atomic units):
\begin{equation}
\label{molecule_shift}
\Delta E = - \left[\frac{4 \sigma}{\sqrt{3}} \epsilon_s \epsilon_p 
\Gamma_{\rm rel} \frac{Z e^2}{a_0} \right]
\frac{d_e}{ea_0}.
\end{equation}

In strong contrast to Eq.(\ref{atom_shift}), the molecular expression
Eq.(\ref{molecule_shift}) does {\it not} depend explicitly on the 
applied electric field ${\bf F}_{\rm applied}$.  The effective electric field is
simply the coefficient in front of $d_e$.  Applying an external
field in the laboratory is essential, however.  In zero electric
field, the molecule's energy eigenstates are also eigenstates of
parity \cite{bcdiat}:
\begin{equation}
\label{parity}
\|\Omega|, \pm \rangle =\frac{1}{\sqrt{2}} 
\left( |\Omega \rangle \pm |-\Omega \rangle \right),
\end{equation}
but separated by an energy $E_{\rm doub}$ referred to as the $\Omega$-doubling
energy. By convention, the lower (upper) sign of this doublet
is denoted by the letter $e$ ($f$).
In such a state, the electronic spin projection $\sigma$ will lie
along the molecular dipole moment as often as against it, canceling the
energy shift.
However, in a modest electric field ${\bf F}_{\rm applied}$,
the signed values of $\pm \Omega$ are again good quantum numbers,
and the expression Eq.(\ref{molecule_shift}) is relevant.  To overcome $E_{\rm doub}$,
the field must be large compared to a ``critical'' field, at which
the Stark energy overcomes the energy difference between the two
parity eigenstates:
\begin{equation}
\label{critical_field}
F_{\rm crit} = \frac{E_{\rm doub}}{d_M},
\end{equation}
where $d_M$ is the permanent electric dipole moment of the molecule.
For the proposed molecular ion experiment,
it is desirable to make $F_{\rm crit}$ as small as possible,
preferably below 100 V/cm \cite{Cornell}.

\section{Computational method}
\label{sec3}

There remains the question of evaluating the constants $\epsilon_s$
and $\epsilon_p$ for a given molecule of interest.  Fortunately,
it is common practice in molecular structure theory to cast molecular
orbitals as linear combinations of atomic orbitals (LCAO). The atomic
orbitals, in turn, are usefully expressed in terms of a Gaussian
basis set that is carefully constructed to reproduce the energy levels
of the individual atoms.  After decades of development, these methods
are now robust and commercially available in software packages
such as Gaussian \cite{gaussian} and MOLPRO \cite{molpro}. We use MOLPRO 
in the calculations performed here.

\subsection{Molecular electronic structure}
\label{molpro_stuff}

We assume that the molecules are well-approximated
by Hund's angular momentum coupling case (a).  This approximation is
justified by the much greater strength of the spin-orbit interaction
and the exchange splitting (both on the order of $10^3$cm$^{-1}$) 
as compared to rotational energies (on the order of $10$ cm$^{-1}$).  
In this case the relevant quantum numbers of
each electron are its projection of orbital ($\lambda$) and 
spin ($\sigma$) angular momentum on the molecular axis.  (In fact,
strong spin-orbit mixing will impact these quantum numbers slightly,
pushing the molecules toward Hund's case (c).  We will deal with this later).

To make a concrete calculation, each molecular orbital is expanded
in a basis set of Gaussian functions, where a basis element is defined as
\begin{equation}
|b_j \rangle = \sum_i s_i|g_i\rangle.
\label{Gaussian_expansion}
\end{equation}
Here each $|g_i \rangle$ is a Gaussian function centered on either of
the atoms in the molecule.  Such a set of Gaussians is chosen to
optimize the energies of each atom separately and is provided in one
of a set of standard basis sets, referred to below.
With this basis, it is now possible to define an orbital as
\begin{equation}
|\phi_k \rangle = \sum_j c_j|b_j\rangle,
\label{orbital_expansion}
\end{equation}
and a configuration is a product of occupations of orbitals given as
\begin{equation}
\label{config_exapansion}
|\Phi_l\rangle = {\cal A} \prod_k d_k |\phi_k\rangle.
\end{equation}
The $d_k$ can take the values 0,1 or 2. ${\cal A}$ stands for anti-symmetrization.

The Hartree-Fock (HF) procedure minimizes the energy of the ground state configuration, 
hence only one configuration is optimized. This is 
done by minimizing the coefficients in the orbitals and configurations. For concreteness, the
configuration is
\begin{equation}
\label{true_config} |\Phi^{\rm HF}\rangle = \prod_k d_k \sum_j c_{kj} |b_j\rangle,
\end{equation}
and the procedure finds the $d_k$ and $c_{kj}$ that minimize the total electronic 
energy $E$. After this procedure, 
we perform a Multi-Configuration Self Consistent Field (MCSCF) \cite{casscf,casscf2} 
calculation with a Complete Active Space (CAS) within MOLPRO. The MCSCF takes the HF $E$ as 
an initial guess and finds appropriate linear combinations of the configurations 
given in Eq.(\ref{true_config})
that minimize $E$ over $M$ configurations. The $M$ configurations can contain 
many different symmetries and spin states. This wave function is 
\begin{eqnarray}
\nonumber |\Phi^{\rm MCSCF}\rangle &=& \sum_l f_l|\Phi_l\rangle\\
\label{mcscf} &=& \sum_l f_l \prod_k d_{lk}\sum_j c_{kj}|b_j\rangle,
\end{eqnarray}
and the procedure varies $f_l$, $d_{lk}$, and $c_{kj}$ in order to minimize the average $E$. This 
is the time-consuming part of the calculation due to the non linear optimization routine. 
This step is then followed by the Multi-Reference Configuration 
Interaction (MRCI) \cite{mrci1,mrci2} program. A MRCI calculation uses the MCSCF 
configurations and mixes in single and double excitations. The wave function is 
\begin{equation}
\label{mrci}
|\Psi^{\rm MRCI}\rangle = \sum_m h_m |\Phi_m^{\rm MCSCF}\rangle,
\end{equation}
where this procedure finds the $h_m$ that minimize the value of $E$ for a given symmetry, 
subject to the constraint that the $f_l$, $d_{lk}$ and $c_{kj}$ are all fixed.

Therefore, deep inside the MCSCF wave function is the $\sigma$
molecular orbital that mixes the atomic $s$ and $p$ orbitals.
We denote this orbital as $|{\rm mol}, \sigma \rangle$, which
has an explicit expansion into the Gaussian basis functions according to
Eq.(\ref{true_config}). Analogously, the wave function
for the heavy atom (e.g., Hf$^+$ or Pt$^+$) is also computed at
the MCSCF and MRCI levels and contains atomic orbitals $|{\rm atom},s \rangle$
and $|{\rm atom}, p \rangle$ expanded into the {\it same}
Gaussian basis set.  It is now a straightforward to estimate the amplitudes 
$\epsilon_{s,p} = \langle {\rm mol} | {\rm atom}, (s,p) \rangle$.

There are two final, but sometimes decisive, corrections.  First,
the desired configuration does not constitute the entire wave function,
but only a fraction of it, given by its amplitude, e.g. $h_0$
in the MRCI expansion in Eq.(\ref{mrci}). For the molecules considered in 
the proposed experiment, however, we have found
this factor to be within several percent of unity. Alternatively, 
it does play a crucial role in the case of PbO where the factor $h_0$ 
is considerably less than unity for the $(\sigma\sigma^*){}^3\Sigma$ 
metastable state; the state that contributes to the EDM signal.

More significantly, the molecular state identified
in Eq.(\ref{mrci}) may be mixed with others via spin-orbit
couplings.  In the case that one other molecular symmetry is mixed
in, we get
\begin{equation}
|\Psi \rangle = 
\cos \chi |^{2S+1}\Lambda \Sigma, \Omega \rangle + 
\sin \chi |^{2S^{\prime}+1}\Lambda^{\prime} \Sigma^{\prime}, 
\Omega \rangle,
\end{equation}
in terms of a mixing angle $\chi$.
There is no prime on the $\Omega$ since spin-orbit interactions preserve the value of 
$\Omega$. The factor $\cos \chi$ can be quite significant.  For example, in a 
$^2 \Pi$ state (as in PbF \cite{PbF} and the Halides HBr$^+$ \cite{Derevianko} 
and HI$^+$ \cite{Derevianko,Isaev}), the single valence electron would be in a
$\pi$ state and would not contribute at all to an effective
electric field ($\cos \chi =0$).  However, spin-orbit mixing with 
nearby $^2 \Sigma$
states can introduce a $\sigma$ electron, but its effective electric field is reduced
by a factor of $\sin \chi$. This mixing is responsible for the large effective 
field in PbF because it has a very large spin-orbit constant. Finally, we arrive at the
expression used to estimate the effective electric field
in our approximation:
\begin{equation}
\label{real_molecule_shift}
F_{\rm eff} = \frac{\Delta E}{d_e} \approx
 - \left[\frac{4 \sigma}{\sqrt{3}} 
h_0 \cos \chi \langle {\rm mol},\sigma | {\rm atom},s \rangle
\langle {\rm mol},\sigma | {\rm atom},p \rangle
\Gamma_{\rm rel} \frac{Z e}{a_0^2} \right],
\end{equation}
in terms of analytical expressions and concrete features of MOLPRO output.

\subsection{Potential curves}

In the available literature, there is no consensus as to the 
true ground state of PtH$^+$ \cite{Ohanessian,dyall,zurita}; 
only one paper attempts to work out the ground state of HfH$^+$ \cite{Ohanessian}.
The debate arises in PtH$^+$ concerning the ordering
of the $^1\Sigma$ and $^3\Delta$ states. Therefore, we use MOLPRO along with
state-of-the-art basis sets \cite{stuttPtHfHg} to compute several electronic states 
of these molecules in addition to the $^3\Delta$ states of interest.

Diatomic molecules belong to the $C_{2\infty}$ point group; however,
MOLPRO only uses Abelian point groups, necessitating the use of the $C_{2v}$
point group for our calculations.  In this point group, the projection
of electronic angular momentum along the diatomic axis ($\Lambda$) will
transform as the following:  $\Lambda=2n$ will transform as A1 symmetry,
$\Lambda=-2n$ as A2, $\Lambda=2n+1$ as B1 and $\Lambda=-(2n+1)$ as B2. 
As a result, the full $\Delta$ symmetry transforms as A1$\oplus$A2, requiring the
MOLPRO calculations to include both the A1 and A2 symmetries.

The MRCI \cite{mrci1,mrci2} calculations are performed after a 
MCSCF \cite{casscf,casscf2} calculation at various internuclear
separations ($1\AA<r<7\AA$). The term ``active space'' 
describes the set of orbitals being used to construct the 
configurations, as in Eq.(\ref{true_config}).  The active space is increased
in one of two ways.  One is to allow the electrons 
to reside in higher-lying orbitals that are nominally unoccupied.
The second is to promote electrons from previously closed shells
into these same orbitals.
Both of these options allow more parameters with which 
to optimize the total energy, but it comes at the expense of an increased 
computational burden.  

The basis sets and active space sizes are bench-marked against the
dissociation energies of \cite{Ohanessian, dyall,zurita}. Our final choice was the
ECP60MWB basis set of the Stuttgart/Cologne group \cite{stuttgart} for the heavy atom along with 
the aug-cc-pVTZ basis \cite{avtz} for the hydrogen. The MWB bases of the 
Stuttgart/Cologne \cite{stuttgart,stuttPtHfHg} group stand for
neutral atom quasi-relativistic potentials.
To make calculations manageable, these basis sets deal explicitly
only with electrons outside of a 60-electron ``core.''  The effect
of these core electrons is incorporated through an effective core
potential (ECP), which is also provided along with the basis sets.  
The ECP's include angular-averaged (i.e., scalar) relativistic effects.

The active space chosen is of the form (A1,B1,B2,A2)$=$(6,2,2,1) for both 
PtH$^+$ and HfH$^+$. This notation 
means that the first 6 A1, 2 B1, 2 B2 and 1 A2 orbitals above the closed space are active. 
In PtH$^+$ and HfH$^+$, we close off orbitals in (2,1,1,0).
Closed orbitals participate in optimizing the energy subject to the constraint that they 
are always doubly occupied. Thus, they do not participate in correlations.
These calculations resulted in dissociation
energies of $3.14$eV(PtH$^+$) and $2.49$eV(HfH$^+$) compared to previous theoretical 
energies of $2.87$eV and $2.50$eV \cite{Ohanessian}. Dyall \cite{dyall} obtains a 
dissociation energy for PtH$^+$ of $2.00$eV while Zurita \textit{et al.} \cite{zurita} 
obtain $2.22$eV, both of which are significantly smaller than our result. 
Our deeper, and presumably more accurate, dissociation energies are most 
likely due to increased computational efficiency over the past decade.
Calculations on PtH$^+$ were completed in approximately 6 cpu-hours per point 
on a 2.4 GHz processor, while for HfH$^+$ they were completed in approximately 
20 cpu-minutes per point.

In order to further test the basis sets, we performed calculations on Pt and Hf to find 
the ionization energies. We ran a MCSCF/MRCI calculation on Pt, Pt$^+$ and Pt$^{++}$ as 
well as for Hf, Hf$^+$ and Hf$^{++}$. We found the first and second ionization energies 
of Pt to be $65000$cm$^{-1}$ and $143000$cm$^{-1}$ and of Hf to be $49600$cm$^{-1}$ and 
$115500$cm$^{-1}$. They agree with the empirical values \cite{m58,bw92a} to within 
$6000$ cm$^{-1}$.

\begin{figure}[ht]
\resizebox{4in}{!}{\includegraphics{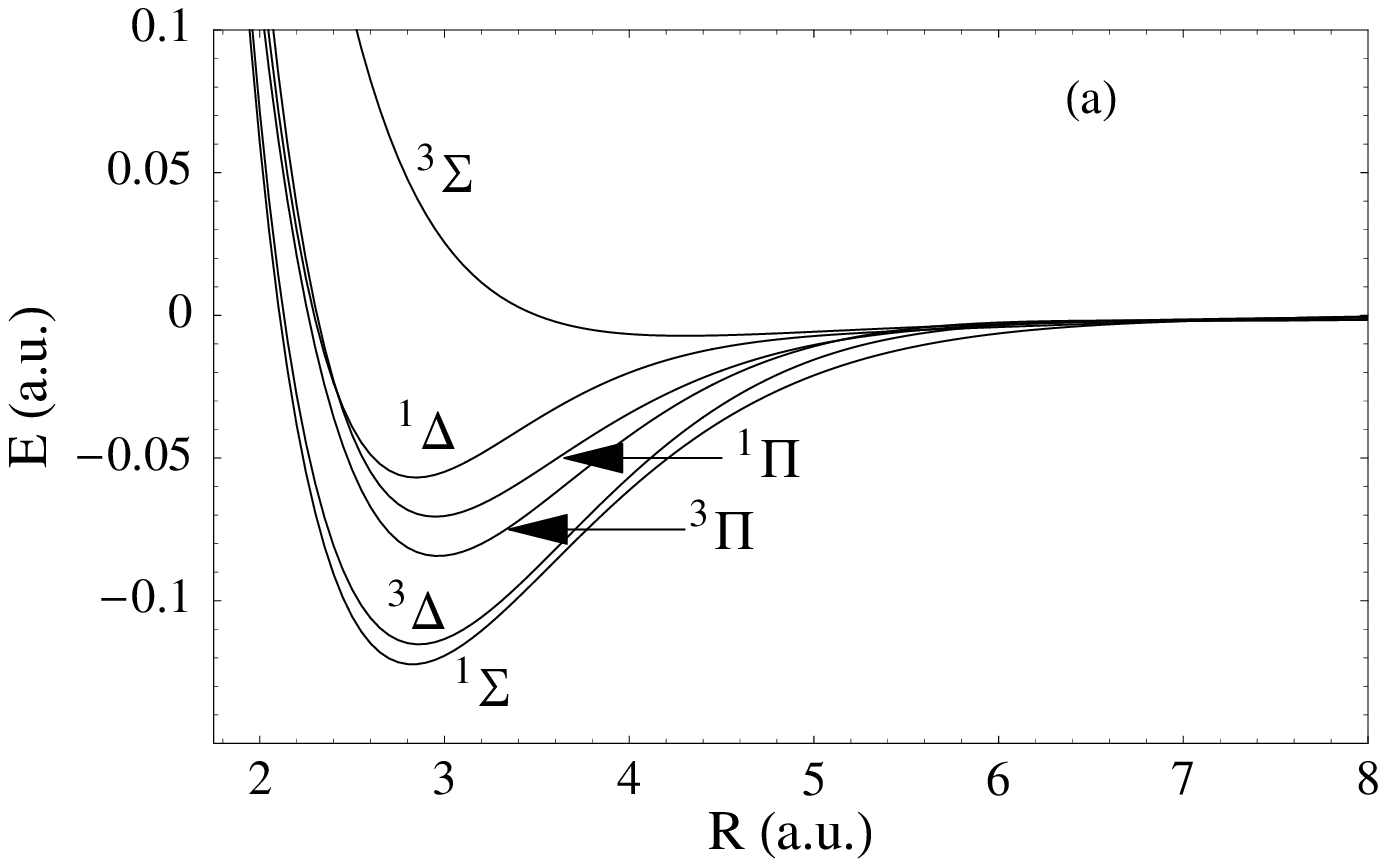}}
\resizebox{4in}{!}{\includegraphics{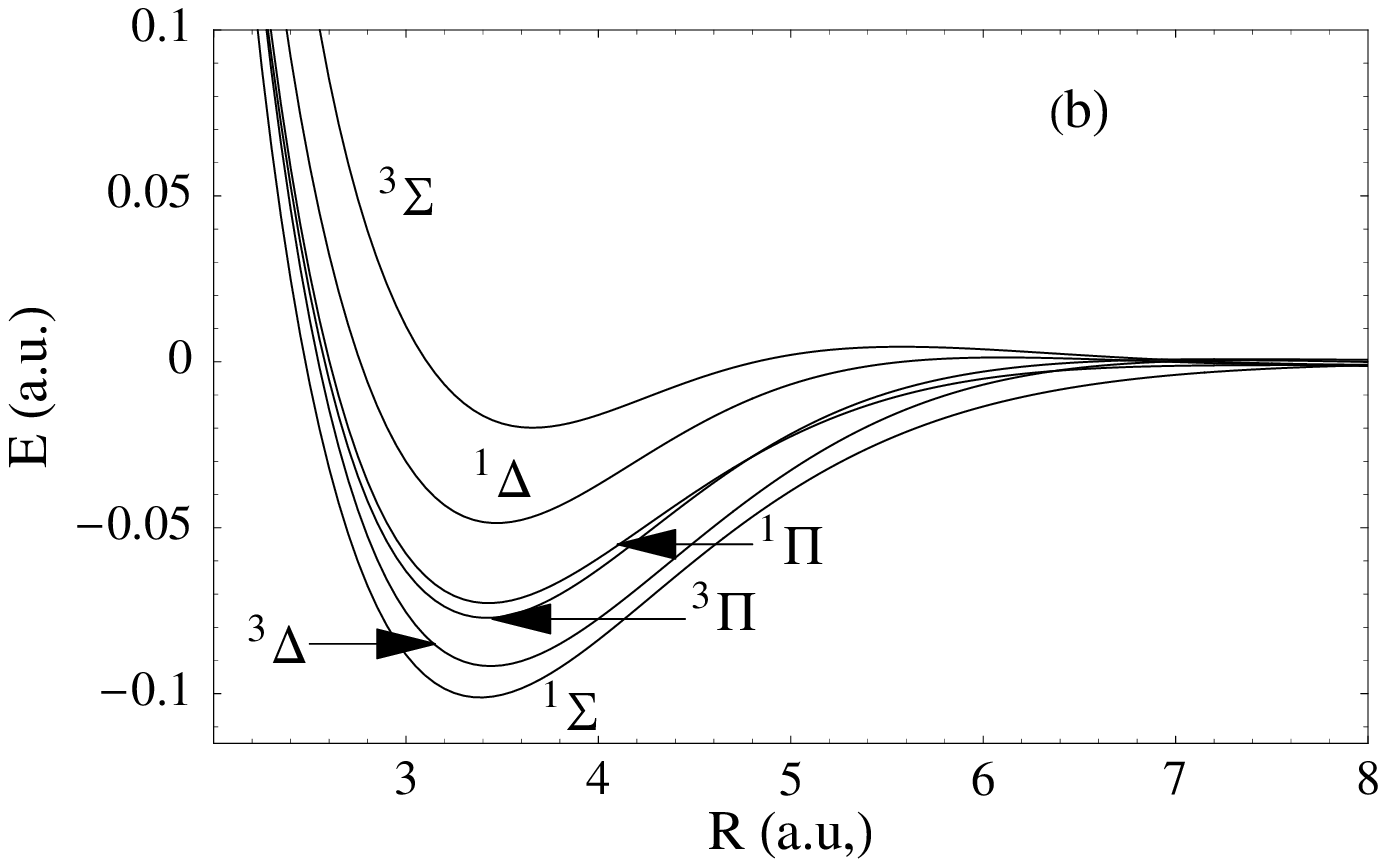}}
\caption{ Potential energy curves for PtH$^+$ (a) and HfH$^+$ (b),
as obtained from MOLPRO output.}
\label{curves_nonrel}
\end{figure}

Born-Oppenheimer potential curves for several symmetries are presented
in Fig.(\ref{curves_nonrel}). The curves are generated by fitting an extended 
Rydberg function to the points obtained from the MOLPRO calculation, using the 
method of Ref.~\cite{aguado}. All curves shown dissociate to
the ground $^2$S state of H and the $^2$D state of the transition
metal ion. To assign
a Hund's case (a) label to a curve, we consult the MOLPRO output and assign 
a value based on the CI eigenvector. In turn, the eigenvector tells in 
which orbitals the electrons lie. There are several different 
configurations present, but the dominant configuration has a 
coefficient $h_0$ very close to unity. As in the case of a PtH$^+$, we find there are 
several configurations, each having two unpaired electrons in the A1 symmetry 
group. Upon examining the orbitals where these electrons lie, we find one has angular 
momentum projection $\lambda=0$, while the other has $\lambda=+2$. Hence, this is
 a $(\sigma\delta)$ 
configuration. Some CI vectors have the same unpaired electrons but have paired 
electrons in a different orbital and/or symmetry group. 
In our approximation, these paired electrons play no role, and 
therefore they contribute nothing to the effective electric field within the molecule. 
Lastly, if this configuration is truly a $^3\Delta$, there should be an 
equivalent energy configuration with one orbital in symmetry A1 and another in 
symmetry A2. This is indeed what we find.

Recall
that Hund's case (a) describes a diatomic molecule with spin and orbital 
motion strongly coupled to the molecular axis. Hence the electronic part of
the Hamiltonian is dominant followed by spin-orbit and then rotational coupling.
Within the case (a) MRCI calculation, both molecules nominally
possess $^1 \Sigma$ ground states.  However, both have low-lying
$^3 \Delta$ states.
The energy difference $E_{^3 \Delta} - E_{^1 \Sigma}$ is 1500 cm$^{-1}$
for PtH$^+$ and  2100 cm$^{-1}$ for HfH$^+$.
At this level we therefore agree qualitatively with the ordering of states given 
by Ref.~\cite{Ohanessian} for PtH$^+$, but not HfH$^+$. 
The equilibrium separations compare
favorably to previous results for PtH$^+$ \cite{Ohanessian,dyall,zurita} and HfH$^+$ 
\cite{Ohanessian}. Rotational constants were not quoted in these papers.
In the $^3\Delta$ states of PtH$^+$ and HfH$^+$ we find 
$r=1.51$\AA, $B_e=7.37$cm$^{-1}$ and $r=1.82$\AA and $B=5.11$cm$^{-1}$.

\subsection{Estimating diagonal spin-orbit contributions}

Roughly speaking, the molecules inherit a large spin-orbit interaction from the heavy ion.  
Moreover, the spin-orbit constant $A$ in PtH$^+$ is expected to be negative, 
as it is in the Pt$^+$ ion, thus shifting the $^3\Delta_3$ state 
lower in energy. For this reason it was previously anticipated 
that $^3\Delta_3$ is likely to be the absolute ground state of PtH$^+$ \cite{dyall}.
By contrast, in HfH$^+$, $A$ is expected to be positive and hence the 
$^3\Delta_1$ state is shifted down in energy. To estimate the size of the shift,
we require knowledge of $A$ in both molecules. Using
methods outlined in Ref.~\cite{bcdiat}, we estimate
the value of $A$ using van Vleck's ``pure precession hypothesis'' \cite{vanvleck}. 
Under this approximation, the spin-orbit coupling arises from the heavy atom, and 
the atomic orbital angular momentum $l$ of each electron is taken to be a good quantum number.

Using this assumption, the ``stretched state'' $^3\Delta_3$ is 
represented by the following Slater determinant:
\begin{equation}
|^3\Delta_3\rangle  =|d\delta^\alpha s\sigma^\alpha|.
\end{equation}
In this expression, we make explicit the orbital angular momentum
quantum numbers $d$ and $s$; the index $\alpha$ stands for
an electron spin aligned along the molecular axis, $\sigma = +1/2.$
The other states can be generated from this one by the application 
of suitable spin-lowering operators.  We can now
act upon wave functions of this type with the single-electron--spin-orbit 
operator, by recasting it in terms of operators acting on the individual electrons $i$:
\begin{eqnarray}
\nonumber H^{LS} &=& A {\bf L} \cdot {\bf S} \\
&=& \sum_i a_i {\bf l}_i \cdot {\bf s}_i.
\end{eqnarray}
Details of this procedure are presented in the Appendix. 
A more thorough treatment of spin-orbit and rotational
coupling will be performed elsewhere \cite{meyer2}.

\begin{figure}[ht]
\resizebox{4in}{!}{\includegraphics{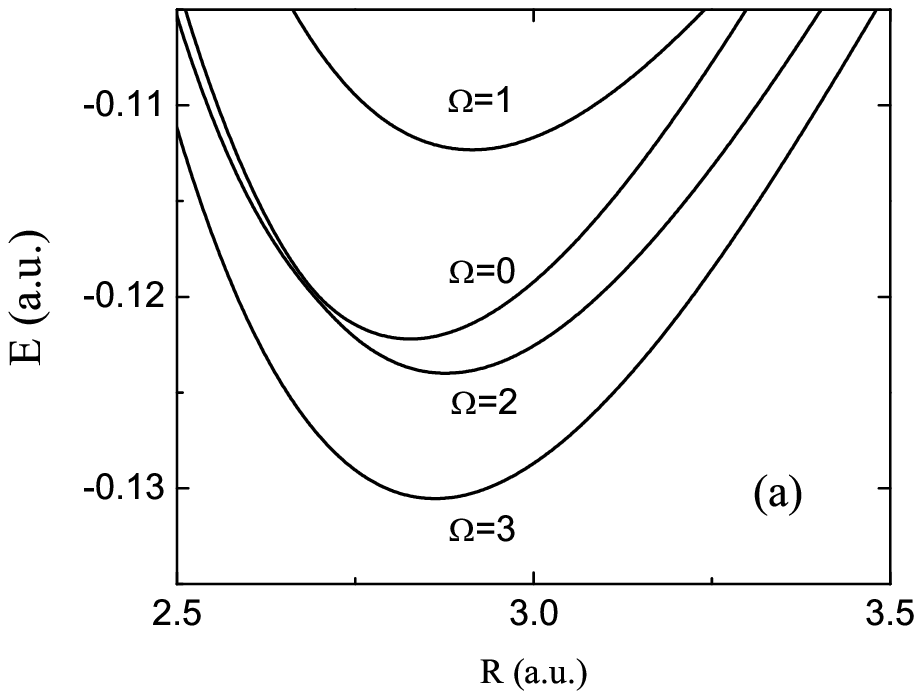}}
\resizebox{4in}{!}{\includegraphics{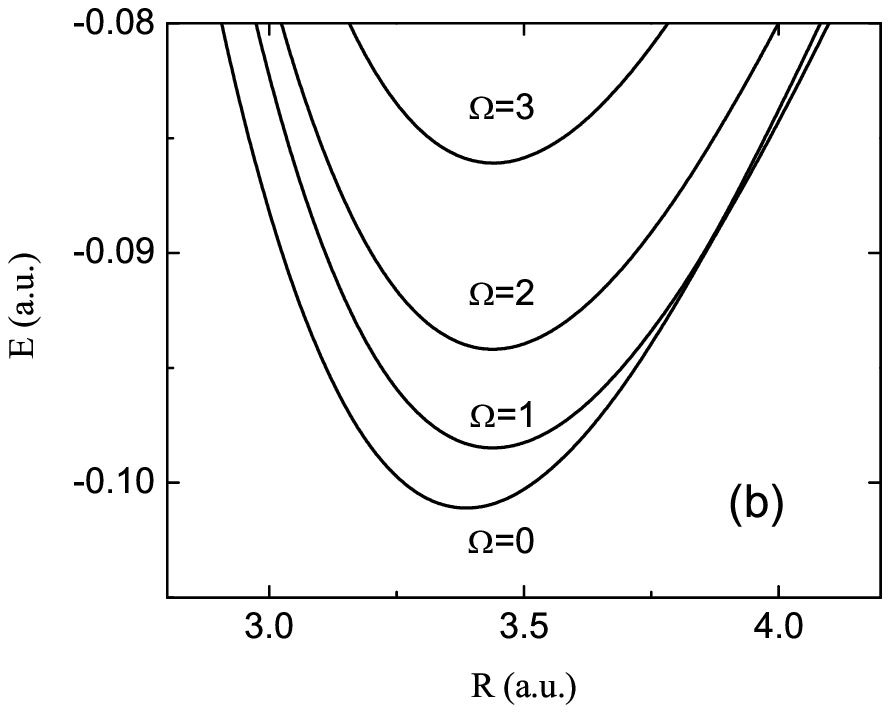}}
\caption{Hund's case (c) potential curves for PtH$^+$ (a) and
HfH$^+$ (b), labeled by the quantum number $\Omega$.  
These curves are obtained from those in Fig.(\ref{curves_nonrel}) by adding
spin-orbit corrections perturbatively, as described in the text.}
\label{curves_rel}
\end{figure}

The effect of the spin-orbit mixing is to break the degeneracy
of the different $\Omega$ values and produce a revised set
of potential curves, indexed by Hund's case (c) quantum numbers.
These curves are shown in Fig.(\ref{curves_rel}).
Performing these calculations yields values of $A=-1680$cm$^{-1}$ and
$A=610$cm$^{-1}$ for the molecular spin-orbit constants of PtH$^+$ and HfH$^+$,
respectively.
This shift is more than enough to push the $^3\Delta_3$ curve below
that of the $^1\Sigma$ in PtH$^+$ (Fig.(\ref{curves_rel}a)). 
Recall that the $^1\Sigma$ state shifts little
in energy because there is no spin or orbital motion, and therefore 
any shift in energy must be a second-order effect. The lowest $\Omega=3$ 
curve is solely $^3\Delta_3$. Additionally, the $^3\Delta_3$ state of PtH$^+$ 
is well described in Hund's case (a); there are no other states with $\Omega=3$ 
nearby in energy (the closest we estimate is a $^3\Phi_3$ state $\approx$ 30,000 cm$^{-1}$
away that dissociates to the $^4$F limit of Pt$^+$).
  
The same results applied to HfH$^+$ tell a different story
(Fig.(\ref{curves_rel}b)).  The value $A$ is
not sufficiently large to shift the $^3\Delta_1$ state below the 
$^1\Sigma_0$ state.  The energy difference between $\Omega=1$ and
$\Omega = 0$ states is 600 cm$^{-1}$. There are other relativistic effects, such as 
the second-order correction to the spin-orbit constant and spin-spin and spin-rotation interactions, 
that can further push down the $^3\Delta_1$ state. For example, 
we can estimate the second-order--spin-spin parameter in terms of the atomic spin-orbit constant 
in the pure-precession approximation \cite{bcdiat}. 
In these heavy elements it is the more dominant contribution to spin-spin. We find
\begin{eqnarray}
\nonumber \lambda^{(2)} &=& -\frac{|\langle ^3\Delta_2|H_so|^1\Delta_2|^2}
{E_{^3\Delta}-E_{^1\Delta}}\\
\nonumber &=& -\frac{1}{2}\frac{a^2}{E_{^3\Delta}-E_{^1\Delta}}\\
\label{spin_spin} &\approx& 60 cm^{-1}
\end{eqnarray}

These contribute to an uncertainty of the an order of magnitude smaller than the 
energy difference in the spin-orbit-corrected curves. It is conceivable that 
the sum total of these other relativistic effects, acting more strongly on the 
$^3\Delta$ state than on the $^1\Sigma$ state, 
will reverse the ordering in Fig.(\ref{curves_rel}b).
In principle, there is an uncertainty in the computed electronic 
energy using MOLPRO, but this is not trivial to estimate. States more 
subject to relativistic effects (like $^3\Delta$) are less accurately 
computed compared to a $^1\Sigma$ state.  More detailed calculations --- 
or, even better, experiments --- will be required to decide this issue.

If it turns out that the ground state of HfH$^+$ is  $^1\Sigma$,
then the $^3\Delta_1$ state is metastable, and an estimate of its
lifetime is desirable. On symmetry grounds we see that this lifetime should be very long, 
because of the fact that the transition would require 
$\Delta S=1$, $\Delta \Omega=1$ and $\Delta \Lambda =2$. ${\bf J}$ is the total 
angular momentum about the internuclear axis, 
${\bf J}={\bf L}+{\bf S}+{\bf N}$, where ${\bf N}$ is the 
mechanical rotation of the nuclei about their center of mass. 
The $S$ and $\Omega$ changes are are not dipole allowed in the case (a) basis. 
However, rotational couplings do not preserve the value of $\Omega$, 
and spin-orbit couplings do not preserve the 
value of $S$.  Because of such interactions, the molecule is not
a pure Hund's case (a) molecule. The $^3\Delta$ states may therefore
by contaminated by small amplitudes of other case (a) states that are
dipole allowed.  These are very small effects and is the reason that these terms 
are usually neglected in the Born-Oppenheimer approximation. 
For example, we can compute the amount of contamination the $^1\Pi_1$ in 
the $^3\Delta_1$ state via spin-orbit. This was done to produce the curves in 
Fig.(\ref{curves_rel}). Using the eigenvectors of the case (c) $\Omega=1$ functions 
we compute the lifetime to be ($|\Omega=1\rangle = \eta |^3\Delta_1\rangle 
+\zeta |^3\Pi_1\rangle+\xi |^1\Pi_1\rangle$):
\begin{eqnarray}
\nonumber \Gamma &\approx& \frac{4}{3}\xi^2\alpha^3(E_{^3\Delta_1}-E_{^1\Sigma_0})^3\\
\nonumber &\sim& 10\;\text{s}.
\end{eqnarray}
We see that lifetime is on the order of seconds.

\subsection{Estimating $\epsilon_s$ and $\epsilon_p$; the effective electric field}

As we have seen, the electric field inside a polar diatomic molecule is 
independent of the applied electric field, provided that the
molecule is polarized. There is a relativistic enhancement
factor determined from properties of the heavy atom and a projection of the 
electron spin, {\bf $\sigma$}, onto the molecular axis. Now we must determine, 
according to our working formula Eq.(\ref{real_molecule_shift}), how much 
the $s$ and $p$ atomic orbitals are mixed because of the presence of the other atom.

\begin{figure}[ht]
\resizebox{4in}{!}{\includegraphics{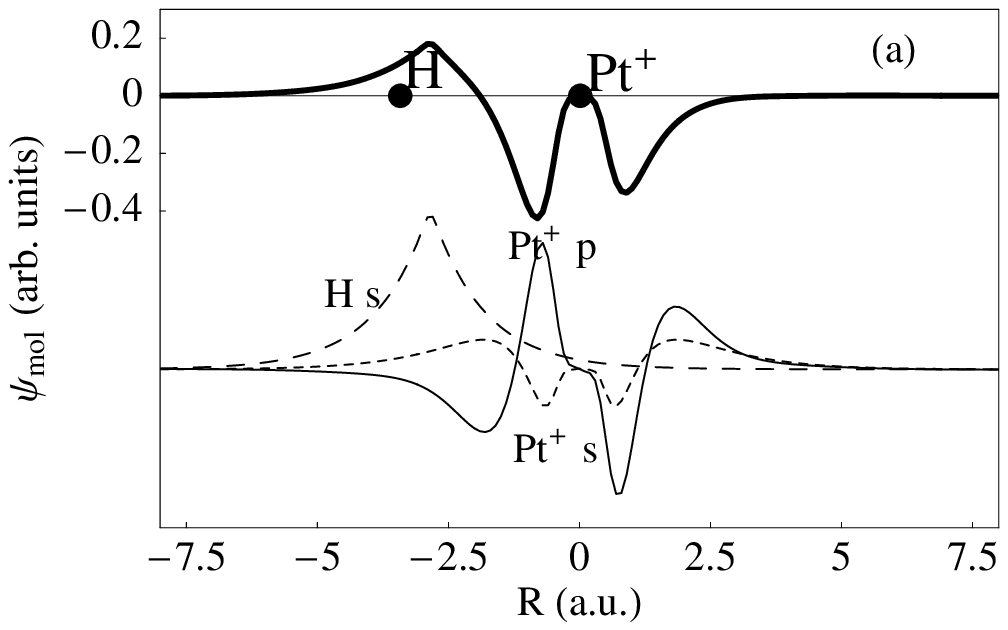}}
\resizebox{4in}{!}{\includegraphics{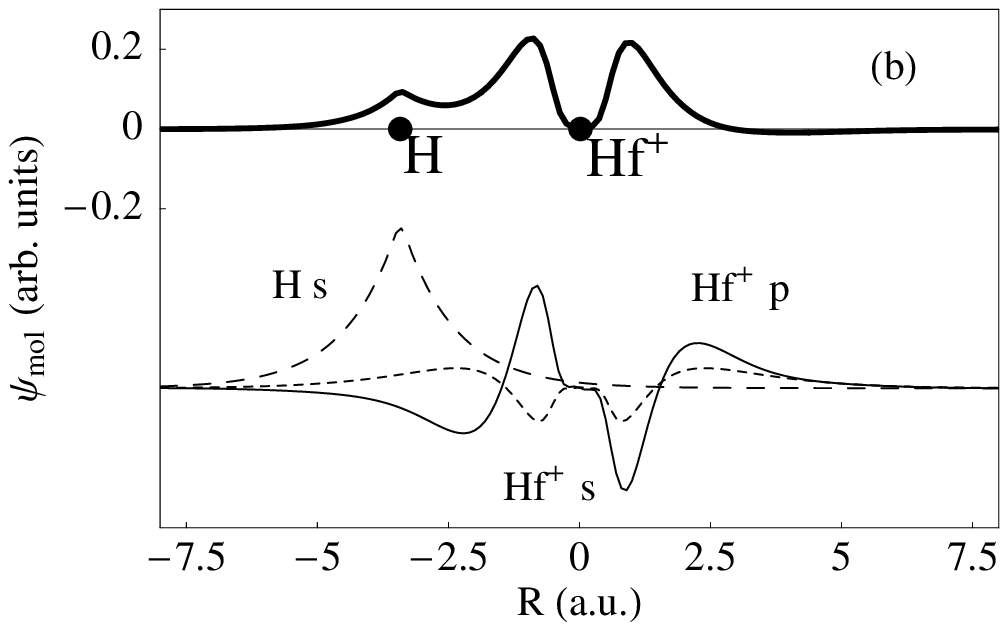}}
\caption{Electronic wave functions (solid lines) for the relevant
$\sigma$ molecular orbitals in the PtH$^+$ (a) and HfH$^+$ (b) ions.
In each figure, the lower panel shows the dominant $s$ and $p$ orbitals
on the heavy ion, plus the $s$ orbital on the hydrogen atom.  All
wave functions are normalized to unity.  By projecting the molecular
orbital onto the atomic orbitals, we find $\epsilon_s = 0.75$,
$\epsilon_p = 0.19$ for PtH$^+$, and $\epsilon_s = 0.79$,  $\epsilon_p= 0.09$
for HfH$^+$.}
\label{wave_functions}
\end{figure}

The values of $\epsilon_s$ and $\epsilon_p$ are obtained from the molecular 
structure calculations, as described above. Wave functions 
generated for the $\sigma$ molecular orbital (not to be confused 
with the electron-spin projection $\sigma$) and presented 
in Fig.(\ref{wave_functions}). In each panel, the electronic
wave function on the molecular axis is plotted as a function of $z$,
the electron's coordinate along this axis (heavy line).  For comparison,
the $s$, $p$, and hydrogen atomic orbitals are also shown below the
molecular orbital wave function.  All wave functions are normalized to
unity.  
Notice that the HfH$^+$ molecular ion in 
Fig.(\ref{wave_functions}b) is not as strongly mixed as in the case of PtH$^+$. 
This small mixing manifests itself as a smaller effective electric field for HfH$^+$ 
(not just due to the $Z$ difference of 78 to 72).
Qualitatively, this
asymmetry illustrates the degree of $s$-$p$ mixing that is vital
to the observed electron EDM shift.  By taking appropriate overlaps of
atomic and molecular wave functions, we construct the effective
electric field as defined in Sec.~\ref{molpro_stuff}.

To check the validity of this procedure we have performed 
the calculations of the expected electric field in molecules for which
${\bf F}_{\rm eff}$ has already been published; e.g., the $^2\Sigma$ fluorides as well 
as the group $VII$ hydrides and PbO in its
metastable $a {}^3\Sigma$ state. Values of $R$ (the internuclear separation) and term 
symbols are taken from the previous work on these molecules. 
Please see Ref.~\cite{BaF} for BaF, Ref.~\cite{PbF} for PbF and HgF, 
Refs.~\cite{Derevianko, Isaev} for HI$^+$, Ref.~\cite{Derevianko} for HrBr$^+$, 
Refs.~\cite{YbF,YbF2} for YbF, and Ref.~\cite{PbO2} for PbO.
We once again used the basis sets of the 
Stuttgart/Cologne \cite{stuttgart} group for the heavy elements. 
Please see Ref.~\cite{stuttBa} for Ba, Ref.~\cite{stuttBr} for Br, 
Ref.~\cite{stuttI} for I, Ref.~\cite{stuttYb} for Yb, Ref.~\cite{stuttPb} 
for Pb and Ref.~\cite{stuttPtHfHg} for Pt, Hf and Hg. The lighter elements H and F use 
the aug-cc-pVTZ basis \cite{avtz}.
These results are presented in 
Table \ref{tab1}.
Although the literature values vary by nearly two orders of magnitude,
our present method tracks most of them to within a factor of two.

The lone exception is HBr$^+$. The sign discrepancy between our value and 
that in Ref. \cite{Derevianko} 
is analogous to the sign difference in HI$^+$, as pointed out in Ref. \cite{Isaev}. 
Differences in the magnitude are likely due to Ref. \cite{Derevianko}'s use of an 
``ionic'' bond, whereas HBr$^+$ is neither 
purely ionic nor covalent. Our method assumes nothing about 
the nature of the bond; rather, it allows the molecular structure 
software to calculate the electron distribution in the molecule and 
then projects the distribution onto atomic orbitals of the heavy atom.

\begin{table}
\caption{\label{tab1} Comparisons of published values of $F_{\rm eff}$ to our 
results using {\it nonrelativistic} software}
\begin{ruledtabular}
\begin{tabular}{l|l|l}
Molecule & Published $F_{\rm eff}$ (GV/cm) & This work $F_{\rm eff}$ (GV/cm)\\
\hline
BaF & 7.4\footnotemark[1] & 5.1\\
\hline
YbF & 26\footnotemark[2] & 43\\
\hline
HgF & 99\footnotemark[3] & 68\\
\hline
PbF & -29\footnotemark[3] & -36.6\\
\hline
PbO & 6.1\footnotemark[4] & 3.2\\
\hline
HBr$^+$ & -0.02\footnotemark[5] & 0.16\\
\hline
HI$^+$ & -0.1\footnotemark[5],0.34\footnotemark[6] & 0.57\\
\hline
PtH$^+$ & N/A & 73\\
\hline
HfH$^+$ & N/A & -17\\
\hline
HfF$^+$ & N/A & -18 
\end{tabular}
\end{ruledtabular}
\footnotetext[1]{Ref. \cite{BaF}.}
\footnotetext[2]{Ref. \cite{YbF,YbF2}.}
\footnotetext[3]{Ref. \cite{PbF}.}
\footnotetext[4]{Ref. \cite{PbO2}.}
\footnotetext[5]{Ref. \cite{Derevianko}, ``Ionic''.}
\footnotetext[6]{Ref. \cite{Isaev}.}
\end{table}

The values we obtain for the effective electric fields are: 
73 GV/cm in the $^3\Delta_3$ state of PtH$^+$ and -17 GV/cm in the 
$^3\Delta_1$ state of HfH$^+$.
Strikingly, these estimates are competitive with the largest ones previously
identified. This result points to the imperative need for a complete
electronic structure calculation to refine the predictions for these
ions.

Finally, at the suggestion of E. Cornell's group, we have used the same methods to 
estimate the effective electric field for HfF$^+$. Like HfH$^+$, its ground state 
is a toss-up between $^3\Delta_1$ and $^1\Sigma_0$, and its effective electric 
field is $F_{\rm eff} \approx -18$  GV/cm.

\subsection{$\Omega$-doubling and critical field}
\label{o-doub}

There remains the question of how easily polarized are the molecular states
we have identified.  
The energy difference between states of differing parity Eq.(\ref{parity}) in a 
$\Delta$ state is a fourth order effect in perturbation theory provided 
the $^{2S+1}\Pi$ and $^{2S+1}\Sigma$ electronic states are fairly well 
separated in energy from the $^{2S+1}\Delta$ state. Brown, \textit{et al.} \cite{brown2}
have used this idea to give formulas for estimating the effect in a manner 
consistent with the ideas of Mulliken and Christy \cite{mull-chris}, who worked
out the effect in $^2\Pi$ states. Thus, the $\Omega$-doubling is relegated to
finding the $\Lambda$-doubling (the difference in energy between parity states comprised of  
$\Lambda=2$ and $\Lambda=-2$).

In a $^3\Delta$ state, there are three parameters contributing to the energy 
splitting designated $\tilde{o}_{\Delta}$, $\tilde{p}_{\Delta}$, and 
$\tilde{q}_{\Delta}$. The 
parameters have the following forms
\begin{eqnarray}
\tilde{o}_\Delta &=& 4 \sqrt{30}\frac{V_A V_A V_B V_B}
{\Delta E_1 \Delta E_2 \Delta E_3} 
\left\{\begin{array}{ccc}
S & S & 2 \\
1 & 1 & S^\prime \end{array}\right\}\frac{(2S-2)!}{(2S+3)!}\\
\tilde{p}_\Delta &=& 4 \sqrt{35}\frac{V_A V_B V_B V_B}
{\Delta E_1 \Delta E_2 \Delta E_3}\\
\tilde{q}_\Delta &=& 8\frac{V_B V_B V_B V_B}
{\Delta E_1 \Delta E_2 \Delta E_3},
\end{eqnarray}
where $V_{A(B)}$ is the matrix element of the spin-orbit (rotational) Hamiltonian 
between differing states. $\langle ^3\Delta_1|A L_-|^3\Pi_1 \rangle$ is an example 
of $V_A$, whereas $\langle ^3\Delta_1|B L_-|^3\Pi_0 \rangle$ is an example of $V_B$.
The $\Delta E_i$ are the energy differences between the state of interest and the 
intermediate states.
It is now straight-forward to estimate the order of magnitude of $E_{doub}$ in the 
$^3\Delta$ states.

In HfH$^+$, we are concerned with the $\Omega=1$ state,
governed by the $\tilde{o}_\Delta$ parameter. We find $E_{doub}$ to be (recall that 
$J=1$ in the ground state)
\begin{eqnarray}
\nonumber
E_{doub}^{\Omega=1} &=& 2 \tilde{o}_\Delta J(J+1)\\
\nonumber &\approx& \frac{\frac{8}{6 \times 5!} \sqrt{30} A^2 B^2}
{(E_{^3\Delta}-E_{^3\Pi})^2 (E_{^3\Delta}-E_{^1\Sigma})} J(J+1)\\
\nonumber &=& 5.1 \times 10^{-5}\;{\rm cm}^{-1}.
\end{eqnarray}
This $E_{\rm doub}$ leads to an $\mathbf{F}_{\rm crit} \approx 1 {\rm V}/{\rm cm}$ 
using the value of $d_m=2.92$ Debye obtained via the MOLPRO calculations.
PtH$^+$ is slightly more complicated. The parameter governing its splitting is 
$\tilde{q}_\Delta$ and connects the $^3\Delta_1$ state to the $^3\Delta_3$ state. 
So, we first find the value of the parameter and then use second-order perturbation 
theory within the $^3\Delta_{\Omega}$ manifold. Thus we find (with $J=3$ in the 
ground state) that
\begin{eqnarray}
\nonumber
E_{doub}^{\Omega=3} &=& \frac{2 q^2 o (J(J+1))^3(J(J+1)-2)(J(J+1)-6)}{(4A-8B)^2}\\
\nonumber &\approx& \frac{4\cdot10^5 B^{10} A^2}
{(4A-8B)^2 (E_{^3\Delta}-E_{^3\Pi})^6(E_{^3\Delta}-E_{^1\Sigma})^3}\\
\nonumber &=& 3.32\times 10^{-20}\;{\rm cm}^{-1}
\end{eqnarray}
Using $d_m=0.77$ Debye in PtH$^+$, from the MOLPRO calculation, we find 
$\mathbf{F}_{crit} \approx 10^{-15} V/cm$, which is negligible 
for any purpose.

\section{Conclusions}

The $^3\Delta$ states of PtH$^+$ and HfH$^+$ satisfy our
two most important criteria for experimental searches of the electron EDM. 
Both are easily polarized and
expected to generate very high effective internal electric fields. 
However, several other experimentally relevant concerns remain to be resolved
for these molecules, including:(1) whether $^3\Delta_1$ is the true
ground state of HfH$^+$; (2) at what laser wavelengths the molecules can
be probed and detected; and (3) how their magnetic properties, e.g.,
g-factors, are influenced by channel couplings engendered by
electric and magnetic fields and terms neglected in the Born-Oppenheimer approximation;
i.e., intermediate coupling via $J^+ S^-$, etc. 
These items demand further detailed study, and will be the subject of future work.

We have endeavored to articulate a strategy for
estimating the influence of the electron EDM on molecular structure,
using perturbation theory and nonrelativistic electronic structure
calculation.  This strategy has proved quite successful and should 
be useful to anyone wishing to assess the viability of a
candidate molecule in the future.  This estimate does not, of course,
take the place of a complete relativistic many-electron calculation,
which must be done as a follow-up.

\acknowledgements

We gratefully acknowledge useful conversations with E. Cornell, A. Leanhardt, L. 
Sinclair, and R. Stutz and thank them for bringing the topic to our attention. 
We are also indebted to L. Gagliardi for discussions on PtH$^+$. ERM thanks 
M. Thompson for some useful discussions pertaining to the MOLPRO software suite for 
heavy atoms. This work was supported by the NSF, the OSEP program at the 
University of Colorado, and the W. M. Keck Foundation.

\appendix

\section{Spin-Orbit in van Vleck Approximation}
First we calculate in terms of Hund's case (a) functions and then in terms of the single 
electron operators. We present the wave functions in terms of Hund's case (a) numbers.
These are presented by $|\Lambda;S,\Sigma;J,\Omega \rangle$ with $J\ge\Omega$. In terms 
of Slater determinants, the molecular wave functions are:
\begin{eqnarray}
\label{del3}|^3\Delta_3\rangle &=& |2;1,1;J,3\rangle\\
\nonumber &=& |d\delta^{\alpha}s\sigma^{\alpha}|\\
\label{del2}|^3\Delta_2\rangle &=& |2;1,0;J,2\rangle\\
\nonumber &=& \frac{1}{\sqrt{2}}(|d\delta^{\alpha}s\sigma^{\beta}|+
|d\delta^{\beta}s\sigma^{\alpha}|)\\
|^3\Delta_1\rangle &=& |2;1,-1;J,1\rangle\\
\nonumber &=& |d\delta^{\beta}s\sigma^{\beta}|\\
\label{pi2}|^3\Pi_2\rangle &=& |1;1,1;J,2\rangle\\
\nonumber &=& |d\pi^{\alpha}s\sigma^{\alpha}|.
\end{eqnarray}
Once again normalizations are assumed in the determinants. The spin-orbit 
energy shift for the $^3\Delta_3$ is then
\begin{eqnarray}
\label{casea}\langle ^3\Delta_3|A L_z S_z|^3\Delta_3\rangle &=& 2A\\
\nonumber \langle|d\delta^{\alpha}s\sigma^{\alpha}||a(l_1^z s_1^z+l_2^z s_2^z)||
d\delta^{\alpha}s\sigma^{\alpha}\rangle &=& a(2(1/2)+0(1/2))\\
\label{onee}&=& a.
\end{eqnarray}
Now we apply the total spin lowering operator 
(given by $S_-=\sum_i s_i^-$) to Eq.(\ref{del3}) to obtain the $^3\Delta_2$ state. 
Since the projection of the spin, $\Sigma$, is zero, there is no shift. Therefore, 
the molecular spin orbital constant $A$ as half the atomic spin-orbit constant $a$. 
The spin-orbit parameter is the same for all states within the $^3\Delta$ manifold.

There are higher-order effects via coupling to other $\Omega=3$ electronic states.
Yet, the only other $\Omega=3$ states come from $^3\Phi$ or $^5\Phi$ states, which lie 
far off in energy. They can be evaluated to see the second-order contribution 
to the molecular spin-orbit constant via a calculation similar to the one below 
describing the coupling of $\Omega=2$ states.

Off-diagonal effects of spin orbit can be evaluated using the same one-electron 
operators. First we write the off diagonal elements using spherical tensor notation.
\begin{equation}
\label{oneeoff}
\frac{a}{2}\sum_{q=\pm1}l_1^{q}s_1^{-q}+l_2^{q}s_2^{-q}.
\end{equation}
Then we evaluate the coupling between Eq.(\ref{del2}) and Eq.(\ref{pi2}). The only 
one electron element that works is $l_1^+s_1^-$ to take $^3\Pi_2$ to $^3\Delta_2$. 
Only including the terms that contribute, we find
\begin{eqnarray}
\nonumber \langle\frac{|d\delta^{\beta}s\sigma^{\alpha}|}{\sqrt{2}}|
\frac{a}{2}l_1^+s_1^-||d\pi^{\alpha}s\sigma^{\alpha}|\rangle &=& \frac{a}{2\sqrt{2}}
\sqrt{(6-2)(3/4+1/4)}\\
&=&\frac{a}{\sqrt{2}}.
\end{eqnarray}

These calculations take into account the permutations of the indices within the 
determinant. This is what cancels the normalization. We have ignored the effects of 
the other electrons in the Slater determinant that belong to paired orbitals. These 
can, in principle, contribute via permutations of the indices and raising/lowering 
the angular momentum. These types of contributions will be evaluated in future 
studies.

\end{document}